\documentclass[12pt]{article}

\usepackage{scicite}

\usepackage{times}

\usepackage{graphicx}
\usepackage{amsmath}  
\usepackage{color}    
\newenvironment{sciabstract}{%
\begin{quote} \bf}
{\end{quote}}

\title{Emergence of Rigidity Percolation in Flowing Granular Systems} 

\author{
Hor Dashti,$^{1,2\ast}$ Abbas Ali Saberi,$^{3,4\dagger}$ S. H. E. Rahbari,$^{2}$ J\"urgen Kurths,$^{5,6}$\\
\\
\normalsize{$^{1}$Australian Institute of Bioengineering and Nanotechnology, }\\
\normalsize{The University of Queensland,
Brisbane, QLD, 4072, Australia}\\
\normalsize{$^{2}$School of Physics, Korea Institute for Advanced Study, Seoul 02455, Korea}\\
\normalsize{$^{3}$Department of Physics, University of Tehran, P. O. Box 14395-547, Tehran, Iran}\\
\normalsize{$^{4}$Max Planck Institute for the Physics of Complex Systems, 01187 Dresden, Germany}\\
\normalsize{$^{5}$Potsdam Institute for Climate Impact Research, Potsdam, Germany }\\
\normalsize{$^{6}$Department of Physics, Humboldt University, Berlin, Germany}\\
\\
\normalsize{$^\ast$E-mail: e.dashti@uq.edu.au} \\
\normalsize{$^\dagger$E-mail: ab.saberi@ut.ac.ir} 
}

\date{}

\begin{document} 

\baselineskip24pt

\maketitle

\begin{sciabstract}

  Jammed granular media and glasses exhibit spatial long-range correlations as a result of mechanical equilibrium. However, the existence of such correlations in the flowing matter, where the mechanical equilibrium is unattainable, has remained elusive. Here, we investigate this problem in the context of the percolation of interparticle forces in flowing granular media. We find that the flow rate introduces an effective long-range correlation, which plays the role of a relevant perturbation giving rise to a spectrum of varying exponents on a critical line as a function of the flow rate. Remarkably, our numerical simulations along with analytical arguments predict a crossover flow rate $\dot{\gamma}_c \simeq 10^{-5}$ below which the effect of induced disorder is weak and the universality of the force chain structure is shown to be given by the standard rigidity percolation. We also find a power-law behavior for the critical exponents with the flow rate $\dot{\gamma}>\dot{\gamma}_c$.

\end{sciabstract}

\section*{Introduction}

One of the principal characteristics of particulate matter, such as granular media and emulsions, is the loss of energy upon every collision, which makes particulate matter an intrinsically non-equilibrium system. However, amorphous particulate materials exhibit many equilibrium-like features, such as the long-range spatial correlations--- a ubiquitous property of equilibrium systems at their critical point. These correlations have been mainly discussed in the context of ({\it i}) interparticle forces, and ({\it ii}) stress components. But understanding how these long-range correlations influence static and dynamic properties of amorphous materials is in its infancy; nonetheless it promises a fascinating avenue of research. This is a subject of fundamental interest for solid mechanics and rheology \cite{Bi}, with applications in resource industries \cite{boger2009rheology}, seismicity \cite{lherminier2019}, material processing, tissue mechanics \cite{bi_2015}, health care, soft robotics and topological metamaterials \cite{mao_2018}. 

In a dense flow regime, unlike its dilute counterpart, grains experience multiple and enduring contacts at all times without the possibility of free flights during an induced shear deformation. This leads the kinematic field of dense granular flows to exhibit complex spatio-temporal correlations and spontaneous formation of quasi-rigid clusters in the flow \cite{radjai2002turbulentlike,rognon2021shear}.
The key to understanding the origins of such correlations is the characterization of interparticle force networks as the backbone of the stability and rigidity of amorphous materials. Percolation theory \cite{saberi_2015}, which describes the connectivity behavior of a network when nodes or links are added, has been widely used for this purpose.

In a seminal work, using photoelastic disks, Majmudar and Behringer \cite{majmudar_2005} found long-range spatial interparticle force correlations for systems subjected to pure shearing and short-range correlations for systems under isotropic compression. This work has been ensued by many experimental and numerical studies using percolation theory for the network of interparticle forces, in which a bond is attributed between two adjacent particles if their interparticle force exceeds a threshold, i.e., $f \ge f_t$. These studies can be divided into two main categories: ({\it  i}) Ostojic et al. \cite{ostojic_2006} found long-range correlations of interparticle forces; Kovalcinova et al. \cite{kovalcinova_2016} also found long-range correlations but with critical exponents not consistent with the previous study in Ref. \cite{ostojic_2006}. Tong et al. \cite{tong_2020} investigated percolation of background forces contributing to mechanical equilibrium in a glass model and found scaling exponents other than the random percolation universality class. In stark contrast to these studies, in category ({\it  ii}), Pathak et al. \cite{pathak_2017} performed a similar numerical analysis on isotropically compressed spheres in two and three dimensions,
and found short range correlations consistent with the random percolation universality class. Accordingly, long-range correlation with debated exponents are found in category-{\it i}, and short-range correlation consistent with the standard random percolation universality class is reported in category-{\it ii}.

There is a large body of studies devoted for percolation of contacts between particles \cite{aharonov_1999,arevalo_2010}; the contact network is the asymptotic limit of the interparticle force network when $f_t\to0$. For instance, Shen et al. \cite{shen_2012} found that $2d$ systems under isotropic compression undergo a connectivity percolation at a density $\phi_P\ll\phi_J$ with a correlation exponent different from that of random percolation. More recent studies have applied techniques of complex networks, such as centrality measures,
to investigate the interplay between various types of centralities and local elastic properties \cite{kollmer_2019}.

\begin{figure}
	\centering
	\includegraphics[width=1.\textwidth]{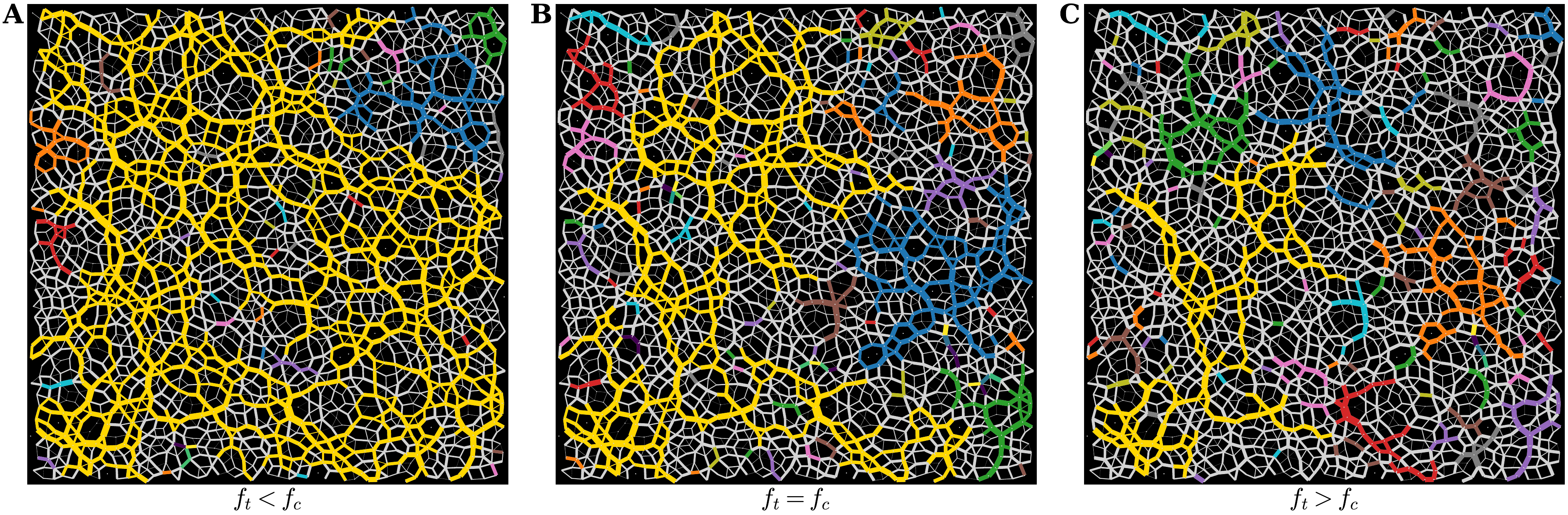}
	\caption{{\bf Snapshots of network of interparticle forces.} Each particle is regarded as a node, and the corresponding interparticle force must be larger than a threshold force, $f>f_t$, to grant a link. By increasing the threshold force, $f_t$, weak interactions are diluted and the largest cluster in the system undergoes a percolation transition at $f_t=f_c$. We identify clusters using a union-find clustering algorithm \cite{knuth2014art}, adopted for off-lattice simulations. In these snapshots, the width of a link is proportional to the strength of the interparticle force. The color coding is performed according to the following rules: diluted weak interactions, for $f<f_t$, are depicted by gray color, the largest cluster is yellow, and the rest of clusters are marked by various other colors. Three different snapshots are shown for (A) $f_t<f_c$, (B) the network is shown at the onset of the percolation transition $f_t=f_c$, and (C) $f_t>f_c$. 
		In these snapshots, the number of particles is $N = 2048$, the packing fraction is $\phi = 0.86$ and $ \dot{\gamma}=10^{-6} $, the threshold force in panel-A to -C, is $f_t = 0.00976$, $0.01084$ and $0.01193$, respectively.}
	\label{fig:snap}
\end{figure}

Whereas the connectivity percolation examines the possibility of a spanning cluster, a more stringent condition is required for rigidity percolation (RP): the spanning cluster must be mechanically rigid. The rigidity percolation was originally proposed to describe the emergence of solidity of covalent network glasses \cite{jacobs_1995}. In athermal systems, rigidity is commonly explored by Hessian, in which the absence of system-spanning zero-cost modes infers rigidity \cite{silbert_2005}. Recently, RP has found many modern applications in phase transitions associated with rigidity in, as diverse systems as, mechanical topological metamaterials \cite{mao_2018}, protein folding \cite{rader_2002}, jamming by compression \cite{ellenbroek_2015}, gelation via attractive interactions \cite{zhang_2019}, and a generalized RP for frictional particles \cite{liu_2019,vinutha_2019}. The universality class of RP is determined by a graph-theoretic technique known as the pebble game \cite{jacobs_1995}. RP is a second ordered phase transition for the bond-diluted generic triangular lattice and the exponents $\nu_{\text{RP}} = 1.21\pm 0.06$, $\beta_{\text{RP}} = 0.18\pm 0.02$ can be obtained by using the cluster moment definitions \cite{jacobs_1995}. Providing the correctness of the hyperscaling relations $ \gamma = d \nu - 2 \beta $, and $ \eta = 2(1 + \beta/\nu) - d $ \cite{stauffer1992intro}, one can calculate the two other important exponents $ \gamma_{\text{RP}} = 2.06\pm 0.08 $, and $ \eta_{\text{RP}} = 0.30\pm 0.02 $. To the best of our knowledge, scaling properties given by the standard RP have never been retrieved in off-lattice simulations of frictionless spheres. Moreover, many studies have investigated a possible connection between jamming in sphere packings and RP, yet, it is shown that RP and jamming are distinct \cite{ellenbroek_2015}.

Here, we report that, at the low limit of the flow (shear) rate below a certain crossover value $\dot{\gamma}_c$, the scaling properties of the percolation network of interparticle forces in particulate matter comply with those of standard rigidity percolation universality class. We argue that the flow rate acts as a relevant perturbation resulting in a spectrum of varying exponents on a critical line. These results shed light on the controversy of the nature of percolation in isotropically compressed packings, because the compression rate may similarly act as a relevant perturbation, and this would explain the range of different exponents reported for the percolation transition in isotropically jammed packings \cite{ostojic_2006,kovalcinova_2016,tong_2020,aharonov_1999,arevalo_2010}. Above $\dot{\gamma}_c$, the induced disorder by the flow rate is long-range enough to drive the universality class of the system with continuously varying critical exponents with $\dot{\gamma}$. However, the critical exponents including the one describing force-force correlation and the fractal dimension of the spanning force cluster, remain unaffected in the whole range of the considered flow rates. Our results should pave the way to improve elasto-plastic models, vastly used in material science and engineering, to include the interplay between the flow rate and force-force correlations as ingredients of the elasto-plastic models. Our study can be suitably extended for understanding of the transitions between different inherent structures and constraint networks in disordered solids \cite{de2009rigidity} when the flow rate is considered as the control parameter. This establishes further analogies between glass and granular physics and their response to external deformations.

\section*{Results}
We perform large-scale molecular-dynamics simulations of two-dimensional \emph{athermal} frictionless bidisperse disks in a simple shear flow using LAMMPS \cite{LAMMPS}. Details on the simulations are given in the Methods section. We build a network of interparticle forces by diluting weak interactions: each particle is considered as a node and a contact is regarded as a link if the interparticle force $f$ exceeds a putative threshold, i.e., $f \geq f_t$. In a jammed state when $f_t$ is small, this network connects most of the particles. By increasing $f_t$, as a result of bond dilution, the largest cluster undergoes a percolation transition at $f_t=f_c$. In Fig. 1-A to -C, we display three snapshots for $f_t<f_c$, $f_t=f_c$, and $f_t>f_c$, respectively. The width of each link is proportional to the magnitude of the interparticle force. The gray color corresponds to diluted weak forces, $f<f_t$. Non-gray colors correspond to existing bonds where $f \ge f_t$. The largest cluster is marked by the yellow color. In panel-B, the largest cluster at the percolation transition contains most of strong forces in the system. 
One can see that the network consists of various types of polygons; triangles seem to have the largest population, in accordance with \cite{arevalo_2010}. In some cases, these triangles act as hinges connecting neighboring polygons. When the threshold is large, panel-C, most of the polygons in the largest cluster are washed out, and chiefly linear structures remain. This emphasizes that the force network consists of two subnetworks: filamentary linear structures carrying most of the stress, and polygonal structures playing the role of stabilizer and hinges for the strong filamentary chains. This is consistent with previous studies \cite{arevalo_2010} and predictions by Radjai et al. \cite{radjai_1996}.
An earlier prediction by Radjai et al. \cite{radjai_1996} stated that the interparticle force network of static jammed materials consists of a subnetwork of strong interactions embedded in another subnetwork of weak interactions. Further investigations have shown that strong forces form linear filamentary chains that are stabilized by weak interactions \cite{arevalo_2010}. One can see in Fig. 1-A that the largest cluster consists of polygons, and neighboring polygons are hinged by triangular structures. 
The formation of such closed loops, which resemble various types of polygons, for frictionless systems is quite surprising, because frictional forces are required to stabilize arches. For this case, an arch is balanced by embedding weak clusters. This is in accord with Arevalo et al. \cite{arevalo_2010}.
At the critical transition point (Fig. 1-B), the yellow cluster contains most of the strong interactions, and when most of the weak interactions are diluted in (Fig. 1-C), the largest cluster contains mostly filamentary chains. Regardless of the value of $f_t$, triangles seem to have the largest population, which is consistent with previous studies \cite{arevalo_2010}: the number of triangles only changes as a function of the coordination number which does not change when $f_t$ is varied. One can see that these topological properties, which were mostly identified for static jammed matter are rather universal, and they govern the structure of flowing matter as well.

\begin{figure}
	\centering
	\includegraphics[width=1.0\textwidth]{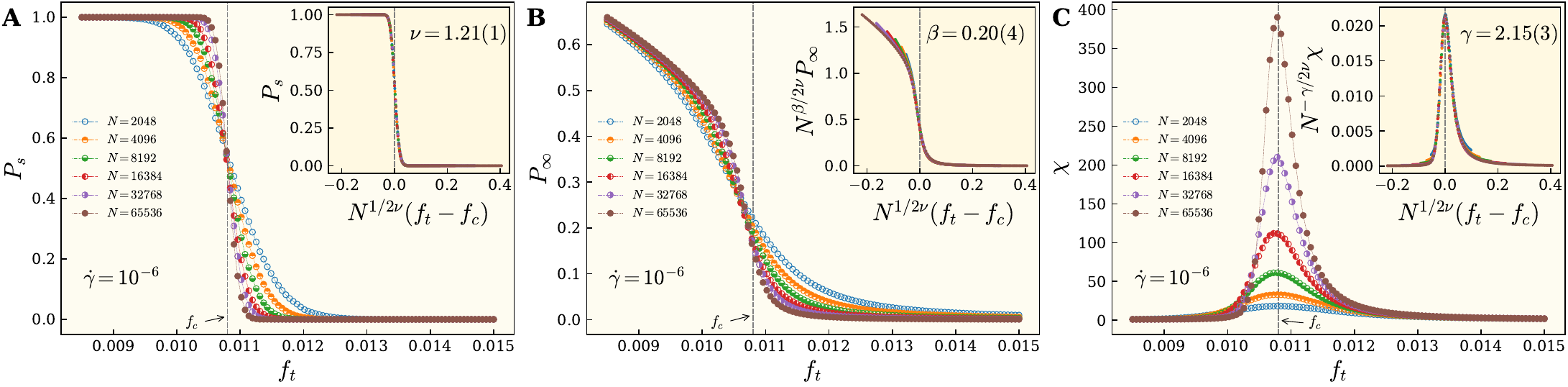}
	\caption[]{{\bf Exponents of the rigidity percolation transition at vanishing rate.} Finite-size scaling is performed at the vanishing limit of flow rate at $\dot{\gamma} = 10^{-6}$, and the packing fraction is $\phi=0.86$. (A) The percolation probability $P_s$ along either side of the system for various sizes is depicted. All curves cross at a common point at $f_c$ signaling the critical nature of the system at the transition point. It is expected that at the infinite system size, $P_s$ becomes a step function. Indeed, as the number of particles is increased, $P_s$ approaches a step function. According to Eq. \eqref{eq:ffs_ps}, by a rescaling according to $N^{1/2\nu} \left(f-f_c\right)$, all curves must collapse into a master function. We obtain an excellent collapse in the inset for $\nu = 1.21 \pm 0.01$. (B) The percolation strength $P_\infty$ is depicted for different sizes. A rescaling according to Eq. \eqref{eq:ffs_pinf}, results to a data collapse at the inset where $\beta = 0.20 \pm 0.04$. (C) Mean cluster size, where the largest cluster is excluded, is depicted for various system sizes. Using Eq. \eqref{eq:ffs_chi}, a scaling collapse is achieved for $\gamma = 2.15 \pm 0.03$ at the inset. A comparison of these exponents with those of the central force rigidity percolation \cite{jacobs_1995} reveals that the universality class obeys the standard RP. Each data point is an average over an ensemble of at least $1.5\times10^4$ configurations. Supplementary Figs. S1-S6 present the details of finite-size scaling and data collapse for various $\dot{\gamma}$.}
	\label{fig:collapse}
\end{figure}

\paragraph*{Exponents of percolation at $\dot{\gamma} \le \dot{\gamma}_c$.} We now characterize the nature of the percolation transition of interparticle forces using scaling analysis of the divergence of correlation length, $\xi \sim |f_t-f_c|^{-\nu}$, the scaling of the percolation strength, $P_\infty \sim |f_t-f_c|^{\beta}$ as the order parameter, and the mean cluster size, $\chi \sim |f_t-f_c|^{-\gamma}$ which acts as the susceptibility in the percolation transition. $\nu$, $\beta$, and $\gamma$ are three critical exponents characterizing the universality class of transition. These exponents can be computed by direct fitting of these functions to the corresponding data. However, in doing so, these exponents will be system size dependent. The universality class of the transition can only be determined, when the values of these exponents are unaffected by the system size. Finite-size scaling is a powerful tool for this purpose, which enables one to extrapolate the exponents corresponding to a system with infinite size. 
We perform systematic finite-size scaling analyses of the percolation probability $P_s$, defined as the fraction of configurations that contain a spanning cluster along $x$- or $y$-direction, the percolation strength $P_\infty$, defined as  the probability that a site belongs to the spanning cluster, as well as susceptibility $\chi$, defined as the mean cluster size excluding the largest cluster. 
We assume the following scaling laws:
\begin{align} 
P_s &= \mathcal{G}_1\left( N^{\frac{1}{d\nu}} \left(f_t-f_c\right) \right) \label{eq:ffs_ps} \\
P_\infty &= N^{\frac{-\beta}{d\nu}}\mathcal{G}_2\left( N^{\frac{1}{d\nu}} \left(f_t-f_c\right) \right) \label{eq:ffs_pinf} \\
\chi &= N^{\frac{\gamma}{d\nu}}\mathcal{G}_3\left( N^{\frac{1}{d\nu}} \left(f_t-f_c\right) \right)
\label{eq:ffs_chi}
\end{align}
where $\mathcal{G}_i$, $i=1$, $2$, and $3$ are scaling functions, $d$ is the dimension of the space ($d=2$ in our case), and $N$ is the number of particles. These equations convey the fact that $P_s$ approaches a step function as $N\to\infty$, $P_\infty$ converges according to $N^{-\beta/d\nu}$, and $\chi$ diverges according to $N^{\gamma/d\nu}$ at $f_t=f_c$. In Fig. 2 panel-A, -B, and -C we show $P_s$, $P_\infty$, and $\chi$, respectively, as a function of a varying force threshold $f_t$ for various system sizes displayed by different colors. The flow rate is $\dot{\gamma} = 10^{-6}$ for all these panels. One can see that $P_s$ approaches a Heaviside function as $N\to\infty$, and all curves corresponding to different system sizes cross at $f_c$. This demonstrates that $P_s$ becomes scale invariant at $f_t=f_c$ signaling the critical nature of the underlying percolation transition. In the panel-A Inset, we obtain an excellent collapse of our data into a master curve after rescaling the horizontal axis according to Eq. \eqref{eq:ffs_ps} with $ N^{1/d\nu} \left( f_t-f_c \right) $, where $\nu = 1.21 \pm 0.01$. The best collapse of $P_\infty$ according to Eq. \eqref{eq:ffs_pinf} in the panel-B Inset suggests $\beta = 0.20\pm0.04$, and similarly we get $\gamma = 2.15 \pm 0.03$ from an excellent collapse of our data according to Eq. \eqref{eq:ffs_chi} in panel-C Inset. 

A comparison of our $\nu$ and $\beta$ exponents with those by the central force rigidity percolation \cite{jacobs_1995} unequivocally establishes that the universality class is the standard rigidity percolation. To the best of our knowledge, this is the first verification of standard RP in an off-lattice molecular dynamics simulation. This paves the way to examine RP in other model glass formers as well as experiments. An ensuing important question is that does the flow rate acts as a relevant parameter in a sense that its variation could drive the universality of the system along a continuous line of classes? In a recent study, it was found that the spatial correlation between bonds in a lattice model for gelation acted as an irrelevant perturbation \cite{zhang_2019}.

\begin{figure}
	\centering
	\includegraphics[width=1.0\textwidth]{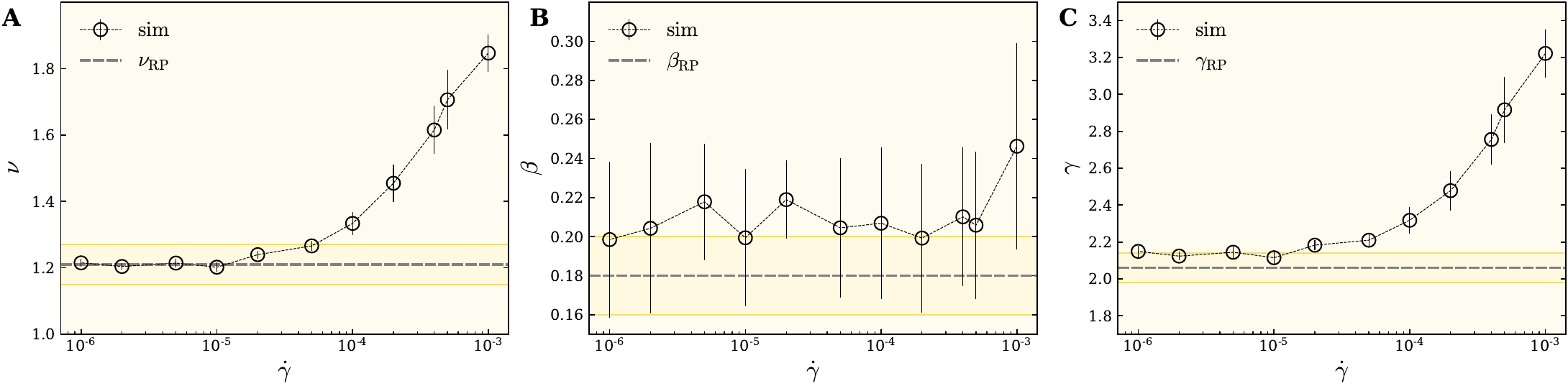}
	\caption{{\bf Exponents of the percolation transition at finite rates.} We vary the flow rate and compute the exponents of percolation transition via finite size scaling for $\nu$, $\beta$ and $\gamma$ exponents in panel-A, -B and -C, respectively. The horizontal dashed line in panel-A, -B and -C show the RP exponents and the shaded area is the corresponding error bar of the exponent. 
		Most of the exponents for $\dot{\gamma} < 10^{-5}$, more or less lie in the standard RP within the error bar. For $\dot{\gamma} > 10^{-5}$, all exponents depart from RP. The packing fraction is $\phi=0.86$, and each data point is an average over an ensemble of at least $1.5\times10^4$ configurations. In Supplementary materials, details of the estimation of error bars of exponents are given. Supplementary Figs. S1-S6 depict finite-size scaling of the above-mentioned critical exponents for various shear rates, $\dot{\gamma}$, which result in the true estimation of the critical exponents in the infinite size limit $N\to\infty$.
	} 
	\label{fig:expo}
\end{figure}

\paragraph*{Exponents of percolation at $\dot{\gamma} > \dot{\gamma}_c$.} To address whether the flow rate acts as a relevant parameter, we vary the flow rate in a range of three orders of magnitude from $\dot{\gamma} = 10^{-6}$ to $10^{-3}$, and for each flow rate we perform a systematic finite-size scaling analysis similar to the aforementioned procedure, to compute critical exponents as a function of the flow rate. Owing to the very large system sizes simulated here, this is a very demanding task, which requires huge processing power and memory storage. We display $\nu$, $\beta$, and $\gamma$ as a function of the flow rate in Fig. 3 panel-A, -B, and -C, respectively, which are estimated in the thermodynamic limit $N\to\infty$. One can see that the flow rate manifests itself as a relevant perturbation and that gives rise to a spectrum of exponents. The horizontal dashed lines in panel-A and -B display RP, and the gray bands show the corresponding uncertainty for RP exponents reported in the existing literature. It is worth noting that for $\dot{\gamma} \le \dot{\gamma}_c\simeq10^{-5}$, all data points remain within the RP band. As a result, for small flow rates, the universality class of the percolation transition remains within RP, but a departure from RP happens at larger flow rates. Yet, the vanishing limit of the flow rate is firmly RP. The correlation exponent $\nu$ increases dramatically for $ \dot{\gamma} > \dot{\gamma}_c \simeq 10^{-5} $. 
Similar observations have been reported for plastic events in sheared amorphous materials, in which at low flow rates plastic events are correlated, yet, at large rates these events become uncorrelated and random \cite{lemaitre_2009,hentschel_2010}.

\paragraph*{Force-force correlations.}
Stress/force correlations have been found in both athermal jammed systems \cite{henkes_2009,nampoothiri_2020,wang_2020_2}, inherent state of supercooled liquids and glasses \cite{charbonneau2014,wu_2015_2,chowdhury_2016,maier_2017,lemaitre_2017,lemaitre_2018,tong_2020}, sheared granular media \cite{zheng_2018_2,wang_2020_2}, and quiescent liquids \cite{lemaitre_2017}. 
A typical stress correlator scales with the distance $r$ as $1/r^d$, where $d$ is the dimension of the space. It is now generally believed that the origin of stress correlation in amorphous materials is the mechanical equilibrium--- Newton's law of force and torque balance. This has been elaborated in recent field-theoretic treatments of amorphous materials \cite{henkes_2009,degiuli_2018_2,lerner_2020}. However, for flowing amorphous material, in which the condition of mechanical equilibrium is broken, the nature of correlations remains elusive. Our percolation framework provides an explicit calculation of some sort of force-force correlations via the scaling of the pair-site correlation function. According to percolation theory \cite{saberi_2015}, two particles separated by a distance $r$, are likely to belong to the same cluster of force chains of strength $f \ge f_c$, by a probability proportional to $g(r)\sim r^{-(d - 2 + \eta)}$. In Fig. 4, we show $\eta$ as a function of the flow rate (see Supplementary Fig. S7 for details of our computations for the correlation function and estimation of the exponent $\eta$ in the infinite size limit $N\to\infty$ for various $\dot{\gamma}$). Interestingly, $\eta$ does not show a systematic dependence in a wide range of flow rates. Moreover, independent of the flow rate, the correlation exponent is given by that of RP (complete results are presented in Supplementary Figs. S1-S8).

\begin{figure}
	\centering
	\includegraphics[width=0.7\textwidth]{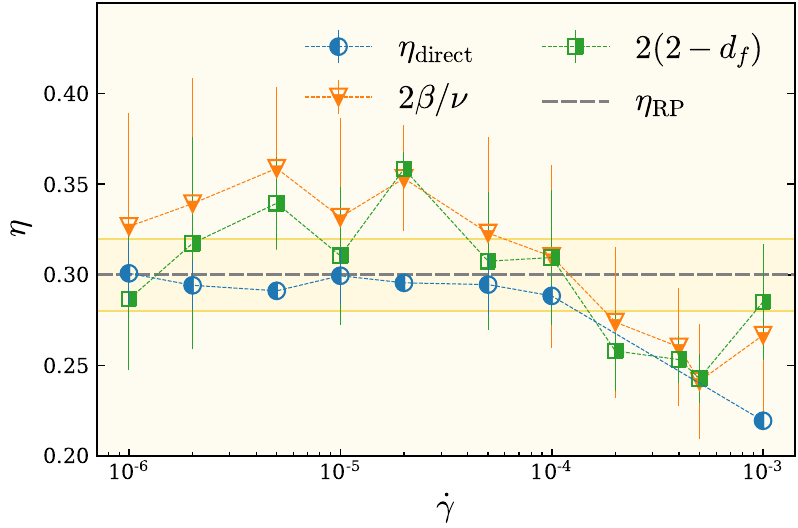}
	\caption{{\bf Exponent of force-force cluster correlation.} 
		$\eta$ is the exponent of force-force cluster correlation calculated via the scaling of pair correlation function $ g(r) \sim r^{-(d - 2 + \eta_{\text{direct}})} $ (circle), and derived from the hyperscaling relation $ \eta=2\beta / \nu $ (triangle) at $f_t=f_c$ (see also Supplementary Fig. S7). 
		The horizontal dashed line shows the RP exponent and the shaded area is the corresponding error bar of the exponent estimated in previous works. Interestingly, $\eta$ does not display a systematic dependence in the wide range of flow rate $\dot{\gamma}$. It is shown that the nature of force-force cluster correlation is given by that of RP for all flow rates. The packing fraction is $\phi=0.86$, and each data point is an average over an ensemble of at least $1.5\times10^4$ configurations.}
	\label{fig:eta}
\end{figure}

\paragraph*{Theoretical framework.}

Our findings suggest that the flow rate $\dot{\gamma}$, as the only parameter in our problem, plays indeed the role of a relevant operator which introduces long-range correlated disorder in our model and can alter the universality of the pure system at $\dot{\gamma} \to 0$. To elucidate our proposition, we draw upon the theoretical framework articulated by the Harris criterion \cite{harris1974effect}. According to this criterion, short-range correlations with a fall-off faster than $r^{-d}$, are relevant if 
$d \nu-2 <0$, where $\nu$ is the correlation length exponent of the pure model, i.e., that of the rigidity percolation at $\dot{\gamma} \le \dot{\gamma}_c$ in our system with $\nu \simeq 1.21$ in two dimensions. Since we have $d\nu-2>0$, so the effective correlations induced by the flow rate in our system can not be short-range but essentially long-range.
For the induced long-range correlations of the power-law form $\mathcal{C}(r)\sim r^{-2H}$ with $2H < d$, the extended Harris criterion \cite{weinrib1984long} then predicts that the correlations are relevant if $H\nu-1<0$. This gives the new correlation length exponent by the scaling relation $\nu_H = 1/H$. This relation has been extensively
verified numerically in various studies in the past \cite{makse2000tracer,marinov2006percolation,prakash1992structural,abete2004percolation,saberi2010geometrical}. In the context of self-affine surfaces (related to our discussion here after we make an analogy with our model for a realization of force profile), it is shown \cite{schmittbuhl1993percolation} that for $H<0$ the percolation is not critical even in the thermodynamic limit and the self-averaging breaks down. For long-range
correlated force configurations with $0\le H< 1$ ($d=2$), in contrast, the transition is critical and the self-averaging is recovered which is in agreement with our observation for various flow rates in this study. The correlation length exponent in the latter case is then shown to be given as follows in terms of the value of $H$:   
\begin{align}
\nu_H =\left\{ \begin{array}{rcl}
1/H  &  \text{if} &  0 < H < 1/\nu,\\
\nu  &  \text{if} &  H \ge 1/\nu.\\
\end{array}\right.
\end{align}
Although this was originally shown for the random percolation model, we note that the same scenario is running here too. According to our results for the exponents presented in Fig. 3-A, for small flow rates of roughly $\dot{\gamma} \leq 10^{-5}$, we find the same correlation exponent $\nu_H \simeq 1.21$ as for the standard rigidity percolation, for which the decay of force-force correlations is governed by the value $H\geq 0.825\pm 0.01$. In the limiting case $\dot{\gamma} \ll \dot{\gamma}_c$ (where $H\to 1$) the correlations will be given by the marginal case proportional to $\propto r^{-d}$ in $d=2$. As the flow rate increases beyond the value $\dot{\gamma}_c \simeq 10^{-5}$, the correlations induced by the flow rate become long-range enough to change the correlation length exponent towards higher values. In particular, we find that for $\dot{\gamma} \simeq 10^{-4}$ the critical exponents are in agreement with the random percolation universality class with $\nu=4/3$; thus giving a proper suggestion $H=3/4$ for the decay of the correlation function, which is in agreement with the previous results in \cite{schrenk2013percolation}. In the infinite flow rate limit $\dot{\gamma} \to \infty$, our theoretical arguments suggest that $H \to 0$, i.e., logarithmic correlation functions appear which are the characteristic feature of turbulence  as a strongly
fluctuating systems in two dimensions \cite{bernard2007inverse}. The best fit to our data for the critical exponent $\nu$ provides the following relation with the flow rate, consistent with our above arguments:
\begin{align}\label{eq:nu-nu_RP}
\nu-\nu_{\text{RP}} = \left\{ \begin{array}{rcl}
a \tilde{\dot{\gamma}}^b \hspace{0.8cm}  &  \text{if} &  \dot{\gamma} > \dot{\gamma}_c\\
0 \hspace{1.1cm} & \text{if} & \dot{\gamma} \le \dot{\gamma}_c,\\
\end{array} \right.
\end{align}
where $\tilde{\dot{\gamma}}=(\dot{\gamma}-\dot{\gamma}_c)/\dot{\gamma}_c$, and $\nu_\text{RP}=1.21$ denotes the exponent for the rigidity percolation, $a=0.027(4)$ and $b=0.73(4)$ (Fig. 5). The crossover flow rate $\dot{\gamma}_c\simeq10^{-5}$ is roughly the point where the system crosses over from a rigid state to the rapid flow regime.

\begin{figure}
	\centering
	\includegraphics[width=0.7\textwidth]{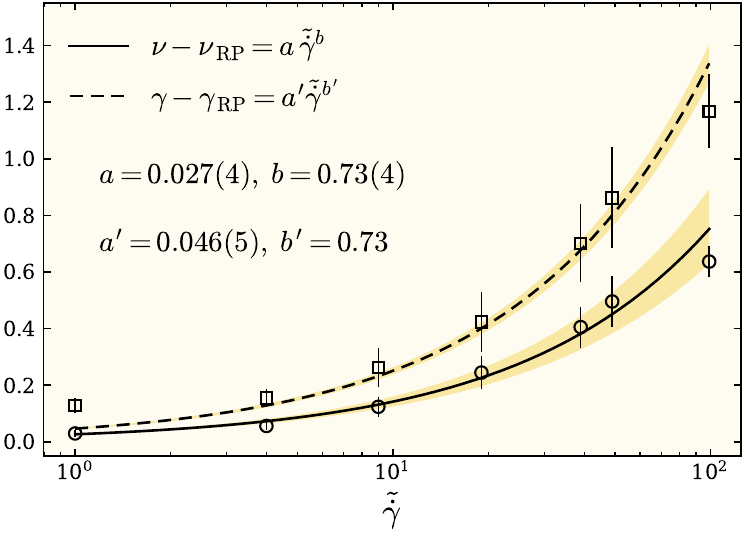}
	\caption{{\bf Power-law scaling of critical exponents.} Above a crossover flow rate $\dot{\gamma}_c\simeq10^{-5}$, the critical exponents $\nu$ and $\gamma$ increase algebraically with the reduced flow rate $\tilde{\dot{\gamma}} = (\dot{\gamma} - \dot{\gamma}_c)/ \dot{\gamma}_c$. Both exponents share the same scaling relation $\sim \tilde{\dot{\gamma}}^{0.73}$ within the error bars. The shaded regions denote the standard error in the fit exponent and the amplitudes over the baselines. $a$ and $b$ are obtained from the best power-law fit to our data for the correlation exponent $\nu$, $b'=b=0.73$ and $a'$ is considered as the only fit parameter for $\dot{\gamma}$. The packing fraction is $\phi=0.86$, and each data point is an average over an ensemble of at least $1.5\times10^4$ configurations.
	}
	\label{fig:nu-gamma}
\end{figure}

In order to further evaluate the geometric response of the force chain structure to the flow rate, we have also measured the fractal dimension $d_f$ of the largest force cluster at $f=f_c$ as a function of $\dot{\gamma}$. The fractal dimension can be estimated from the scaling relation $M \sim N^{d_f/2}$, with $M$ being the average number of nodes that belong to the percolating force cluster. Our results are presented in Fig. 4 in the form $2(2-d_f)$ to further test if it fulfills the hyperscaling relation $\eta = 2+d-2d_f$ (see Supplementary Fig. S8 for more details). As shown in Fig. 4, this scaling relation seems to be valid within the error bars, with an additional observation that the exponents in Fig. 4 show a weak dependence over the whole range of considered $\dot{\gamma}$. To assess the validity of this observation, we take advantage of the scaling relation $\eta=2-\gamma/\nu$ according to which, in order for $\eta$ to be independent of $\dot{\gamma}$, the dependence of $\gamma$ and $\nu$ must be canceled out. According to the relation \eqref{eq:nu-nu_RP}, this requires $\gamma-\gamma_{RP}=a'\tilde{\dot{\gamma}}^{b'}$ with $b'=b=0.73$. Fig. 5 shows our best fit $a'\tilde{\dot{\gamma}}^{0.73}$ to the data for $\gamma-\gamma_{RP}$ as a function of $\tilde{\dot{\gamma}}$ for $\dot{\gamma} > \dot{\gamma}_c$, which gives $a'=0.046(5)$. Remarkably this gives back $\eta=2 - a'/a \simeq 0.30$ which is in perfect agreement with our observation shown in Fig. 4. It must basically be possible to check similar behavior for the exponent $\beta$, but since our data presented in Fig. 3-B bear rather large error bars, our test was not conclusive enough to be shown in Fig. 5.

\section*{Discussion}

The interplay between  network topology and macroscopic properties in network glass formers \cite{he_1985}, and more specifically  in chalcogenide network glasses \cite{narayanan_1996,uebbing_1996}, has been well addressed in the context of rigidity percolation \cite{jacobs_1995}. Attempts have been made to reconcile RP with isotropic jamming \cite{ellenbroek_2015}, yet, such a connection has been evasive. The reconciliation of RP with jammed granular media is crucial because such a reconciliation will facilitate a unified framework for granular media and glasses, as suggested by long-range stress correlations in both systems. 

Here, we investigated the percolation of interparticle forces between particles in a dense regime in which all degrees of freedom of particles are over-constrained \cite{phillips_1979}. We unequivocally established that the connectivity of the interparticle force network undergoes a second-order phase transition, whose scaling properties, for a wide range of flow rates, are given by those of the rigidity percolation universality class. This demonstrates that the emergence of shear elasticity originates from the internal organization of particles  in the form of chains of inter-particle forces. Thus, the emergence of shear elasticity is a manifestation of mechanical self-organization. Recent work supports our view; e.g., Tong et al. \cite{tong_2020} investigated stress correlation in a model glass at finite temperature with giant anharmonic fluctuations, and demonstrated that the stress correlation emerges as a result of  long-range interparticle force correlation manifested as a distinct  correlated percolation universality class--- a hint to mechanical self-organization. 

A crucial step towards a better understanding of force/stress anomalies requires lifting the constraint of the mechanical balance. This is very important for  parent liquid states and flowing granular media that are not quenched into a local energy minimum. As pointed out by Lemaitre \cite{lemaitre_2017}, this is very challenging because, in the absence of the mechanical equilibrium, the structure of correlations becomes very complex. Our work is a step forward in this direction. It provides a previously unknown link between force correlation in flowing granular media, rigidity percolation, and a potential relationship to mechanical self-organization. The ingredients giving rise to RP in sheared frictionless spheres, namely ({\it i}) over-constraining and ({\it ii}) slow deformation, are important and may motivate future attempts for field-theoretic treatments \cite{henkes_2009,degiuli_2018_2} of amorphous flowing material.

Our results cover a typical athermal system undergoing a shear-driven jamming transition in the dense phase. An important question is whether and to what extent these results apply to thermal glasses. The rigidity percolation has been originally devised to address the glass transition in thermal glasses, yet it has not been confirmed in any molecular simulation. Determining whether a slow deformation is a necessary factor for recovering rigidity percolation in thermal glasses will be an exciting research pursuit for future numerical investigations. We hope our work will inspire further studies on thermal glasses using molecular dynamics simulations.

A recent study on jamming by compression shows that the compression rate is a relevant parameter for the jamming transition \cite{peshkov_2021}. Moreover, the authors find that isotropically compression-driven  jamming is in the same universality class as that of  shear-driven jamming. Another study shows that whereas jamming by compression is dramatically history-dependent and as a result gives rise to a range of critical densities, shear-driven jamming is not history-dependent \cite{jin_2021}. There is a chance that the controversy in percolation in isotropically compressed packings is related to these issues and the history of the system plays the role of a relevant parameter giving rise to a spectrum of critical exponents. Our results should motivate future works to resolve this controversy.

\section*{Methods}
Our system consists of frictionless bidisperse disks in two-dimensional space. The interactions between disks are governed by short-range linear repulsive and dissipative forces. These interactions can be expressed as ${\bf F}_{ij}= K_n \xi_{ij} {\bf r}_{ij}/r_{ij} - M_{\mathrm{eff}} \gamma_n {\bf v}_n$.
Two particles $i$ and $j$ of radii $R_i$ and $R_j$ at positions ${\bf r}_i$ and ${\bf r}_j$ interact when the mutual compression of particles $\xi_{ij} = R_i + R_j - r_{ij} > 0$. In this equation, ${\bf r}_{ij} = {\bf r}_i - {\bf r}_j$, $K_n$ is the elastic constant for a normal contact, and $\gamma_n$ is the viscoelastic damping constant for a normal contact.
${\bf v}_n$ is the normal component of the relative velocity of the two particles, and $M_{\mathrm{eff}}= m_1 m_2 / (m_1 + m_2)$. 
The $\xi_{ij}$ is called the mutual compression of two particles.
We utilize LAMMPS software \cite{LAMMPS} for simulating shear-driven granular systems. The Verlet method \cite{allen2017computer}, which serves as the default integrator in LAMMPS, is employed to numerically integrate the equations of motion for particles. This algorithm calculates the new positions and velocities of particles using their current values and the forces acting on them by employing a two-step process that accounts for both position and velocity updates. The details of the algorithm are incorporated in Section III of the Supplementary Information. 
Lees-Edwards boundary conditions \cite{allen2017computer} are applied to create a uniform overall flow rate $\dot{\gamma}$ along the $x$-direction. To emulate these boundary conditions in LAMMPS, each particle is given an initial velocity according to $\mathbf{v}_i(t=0) = \dot{\gamma} y_i \hat{\mathbf{i}}$ at time $t=0$, where $y_i$ is the $y$-position of particle $i$. To maintain the velocity profile, we use the command \texttt{"fix deform"} with \texttt{"remap v"} option.

To prevent crystallization, we employ a $1:1$ binary mixture of particles with particle radii $R_0 = 0.5$ and $R_1 = 0.7$. For seeking simplicity, we set the mass of each particle equal to its area, $m = \pi R^2$.
By considering $K_n=1$ and $\gamma_n=1$, we measure time in our simulations using the units of $\tau_0 \equiv \gamma_n d_0^2/ K_n = 1$, where $d_0$ represents the diameter of the small particles. The shear rate can then be nondimensionalized with $\dot\gamma \times \tau_0$, which is important for comparison with other simulations and experiments.
The packing fraction $\phi=0.86>\phi_J$ is considered.
To check the robustness of the results against changing the packing fraction and the repulsive force amplitude, we simulate systems with  $\phi=0.865, K_n=1$, and $\phi=0.86, K_n=1.5$,  as shown in Figs. S9-S13. Notably, we find that these changes do not significantly affect the reported original exponents, as they fall within the same range of error bars (see Fig. S13).

The number of particles is $N = 2048, 4096, 8192, 16384, 32768$, and $65536$. This vast range of system sizes facilitate a systematic finite-size scaling, by which we calculate all the scaling properties of networks at the thermodynamic limit $N\to\infty$. Furthermore, various flow rates $\dot{\gamma} = 10^{-6}, 2 \times 10^{-6}, 5 \times 10^{-6}, 10^{-5}, 2 \times 10^{-5}, 5 \times 10^{-5}, 10^{-4}, 2 \times 10^{-4}, 4 \times 10^{-4}, 5 \times 10^{-4}$, and $10^{-3}$ are considered for each system size.
All reported quantities, such as $P_s$, $P_{\infty}$ and $\chi$ are computed by averaging over $10$ independent simulations (realizations) each of which consists of around $1.5 \times 10^3$ uncorrelated configurations separated by a strain difference equal to the unit of the length after the system has reached a steady state.

\section*{Acknowledgments}
This work is supported by the Center for Advanced Computation at Korea Institute for Advanced Study. A.A.S. acknowledges the support from the Alexander von Humboldt Foundation (DE) and the research council of the University of Tehran. 


\paragraph*{Author contributions:} 
A.A.S. and J.K. proposed and designed the research. H.D., A.A.S., S.H.E.R., and J.K. conceived the study, carried out the analysis, and prepared the manuscript. H.D. and A.A.S. analyzed data. H.D., A.A.S., S.H.E.R., and J.K. discussed the results and contributed to writing the manuscript. H.D. carried out the numerical simulations. S.H.E.R. prepared the initial version of the draft. A.A.S. and J.K. developed the theoretical framework and drafted the final version of the manuscript.

\paragraph*{Competing interests:} All authors declare that they have no competing interests.

\paragraph*{Data and materials availability:} All data needed to evaluate the conclusions in the paper are present in the paper and/or the Supplementary Materials.

\section*{Supplementary materials}
Supplementary Text\\
Figs. S1 to S13\\

\bibliography{refs}

\begin{thebibliography}{10}

\bibitem{Bi}
D.~Bi, J.~Zhang, B.~Chakraborty, R.~P. Behringer, Jamming by shear, {\it
  Nature\/} {\bf 480}, 355 (2011).

\bibitem{boger2009rheology}
D.~V. Boger, Rheology and the resource industries, {\it Chemical Engineering
  Science\/} {\bf 64}, 4525 (2009).

\bibitem{lherminier2019}
S.~Lherminier, {\it et~al.\/}, Continuously sheared granular matter reproduces
  in detail seismicity laws, {\it Physical review letters\/} {\bf 122}, 218501
  (2019).

\bibitem{bi_2015}
D.~Bi, J.~H. Lopez, J.~M. Schwarz, M.~L. Manning, A density-independent
  rigidity transition in biological tissues, {\it Nature Physics\/} {\bf 11},
  1074 (2015).

\bibitem{mao_2018}
X.~Mao, T.~C. Lubensky, Maxwell lattices and topological mechanics, {\it Annual
  Review of Condensed Matter Physics\/} {\bf 9}, 413 (2018).

\bibitem{radjai2002turbulentlike}
F.~Radjai, S.~Roux, Turbulentlike fluctuations in quasistatic flow of granular
  media, {\it Physical review letters\/} {\bf 89}, 064302 (2002).

\bibitem{rognon2021shear}
P.~Rognon, M.~Macaulay, Shear-induced diffusion in dense granular fluids, {\it
  Soft Matter\/} {\bf 17}, 5271 (2021).

\bibitem{saberi_2015}
A.~A. Saberi, Recent advances in percolation theory and its applications, {\it
  Physics Reports\/} {\bf 578}, 1 (2015).

\bibitem{majmudar_2005}
T.~S. Majmudar, R.~P. Behringer, Contact force measurements and stress-induced
  anisotropy in granular materials, {\it Nature\/} {\bf 435}, 1079 (2005).

\bibitem{ostojic_2006}
S.~Ostojic, E.~Somfai, B.~Nienhuis, Scale invariance and universality of force
  networks in static granular matter, {\it Nature\/} {\bf 439}, 828 (2006).

\bibitem{kovalcinova_2016}
L.~Kovalcinova, A.~Goullet, L.~Kondic, Scaling properties of force networks for
  compressed particulate systems, {\it Physical Review E\/} {\bf 93}, 042903
  (2106).

\bibitem{tong_2020}
H.~Tong, S.~Sengupta, H.~Tanaka, Emergent solidity of amorphous materials as a
  consequence of mechanical self-organisation, {\it Nature communications\/}
  {\bf 11}, 1 (2020).

\bibitem{pathak_2017}
S.~N. Pathak, V.~Esposito, A.~Coniglio, M.~P. Ciamarra, Force percolation
  transition of jammed granular systems, {\it Physical Review E\/} {\bf 96},
  042901 (2017).

\bibitem{aharonov_1999}
E.~Aharonov, D.~Sparks, Rigidity phase transition in granular packings, {\it
  Physical Review E\/} {\bf 60}, 6890 (1999).

\bibitem{arevalo_2010}
R.~Arevalo, I.~Zuriguel, D.~Maza, Topology of the force network in the jamming
  transition of an isotropically compressed granular packing, {\it Physical
  Review E\/} {\bf 81}, 041302 (2010).

\bibitem{shen_2012}
T.~Shen, C.~S. O'Hern, M.~D. Shattuck, Contact percolation transition in
  athermal particulate systems, {\it Physical Review E\/} {\bf 85}, 011308
  (2012).

\bibitem{kollmer_2019}
J.~E. Kollmer, K.~E. Daniels, Betweenness centrality as predictor for forces in
  granular packings, {\it Soft Matter\/} {\bf 15}, 1793 (2019).

\bibitem{knuth2014art}
D.~E. Knuth, {\it Art of computer programming, volume 2: Seminumerical
  algorithms\/} (Addison-Wesley Professional, 2014).

\bibitem{jacobs_1995}
D.~J. Jacobs, M.~F. Thorpe, Generic rigidity percolation: the pebble game, {\it
  Physical Review Letters\/} {\bf 75}, 4051 (1995).

\bibitem{silbert_2005}
L.~E. Silbert, A.~J. Liu, S.~R. Nagel, Vibrations and diverging length scales
  near the unjamming transition, {\it Physical Review Letters\/} {\bf 95},
  098301 (2005).

\bibitem{rader_2002}
A.~J. Rader, B.~M. Hespenheide, L.~A. Kuhn, M.~F. Thorpe, Protein unfolding:
  rigidity lost, {\it Proc. Nat. Acad. Sci. USA\/} {\bf 99}, 3540 (2002).

\bibitem{ellenbroek_2015}
W.~G. Ellenbroek, V.~F. Hagh, A.~Kumar, M.~F. Thorpe, M.~Van~Hecke, Rigidity
  loss in disordered systems: Three scenarios, {\it Physical Review Letters\/}
  {\bf 114}, 135501 (2015).

\bibitem{zhang_2019}
S.~Zhang, {\it et~al.\/}, Correlated rigidity percolation and colloidal gels,
  {\it Physical Review Letters\/} {\bf 123}, 058001 (2019).

\bibitem{liu_2019}
K.~Liu, S.~Henkes, J.~M. Schwarz, Frictional Rigidity Percolation: A New
  Universality Class and Its Superuniversal Connections through Minimal
  Rigidity Proliferation, {\it Physical Review X\/} {\bf 9}, 021006 (2019).

\bibitem{vinutha_2019}
H.~A. Vinutha, S.~Sastry, Force networks and jamming in shear-deformed sphere
  packings, {\it Physical Review E\/} {\bf 99}, 012123 (2019).

\bibitem{stauffer1992intro}
D.~Stauffer, A.~Aharony, {\it Introduction to percolation theory\/} (Taylor and
  Francis, 1992).

\bibitem{de2009rigidity}
V.~K. de~Souza, P.~Harrowell, Rigidity percolation and the spatial
  heterogeneity of soft modes in disordered materials, {\it Proceedings of the
  National Academy of Sciences\/} {\bf 106}, 15136 (2009).

\bibitem{LAMMPS}
A.~P. Thompson, {\it et~al.\/}, {LAMMPS} - a flexible simulation tool for
  particle-based materials modeling at the atomic, meso, and continuum scales,
  {\it Comp. Phys. Comm.\/} {\bf 271}, 108171 (2022).

\bibitem{radjai_1996}
F.~Radjai, M.~Jean, J.~J. Moreau, S.~Roux, Force Distributions in Dense
  Two-Dimensional Granular Systems, {\it Physical Review Letters\/} {\bf 77},
  274 (1996).

\bibitem{lemaitre_2009}
A.~Lemaitre, C.~Caroli, Rate-Dependent Avalanche Size in Athermally Sheared
  Amorphous Solids, {\it Physical Review Letters\/} {\bf 103}, 065501 (2009).

\bibitem{hentschel_2010}
H.~G.~E. Hentschel, S.~Karmakar, E.~Lerner, I.~Procaccia, Size of Plastic
  Events in Strained Amorphous Solids at Finite Temperatures, {\it Physical
  Review Letters\/} {\bf 104}, 025501 (2010).

\bibitem{henkes_2009}
S.~Henkes, B.~Chakraborty, Statistical mechanics framework for static granular
  matter, {\it Physical Review E\/} {\bf 79}, 061301 (2009).

\bibitem{nampoothiri_2020}
J.~N. Nampoothiri, {\it et~al.\/}, Emergent elasticity in amorphous solids,
  {\it Physical Review Letters\/} {\bf 125}, 118002 (2020).

\bibitem{wang_2020_2}
Y.~Wang, Y.~Wang, J.~Zhang, Connecting shear localization with the long-range
  correlated polarized stress fields in granular materials, {\it Nature
  communications\/} {\bf 11}, 1 (2020).

\bibitem{charbonneau2014}
P.~Charbonneau, J.~Kurchan, G.~Parisi, P.~Urbani, F.~Zamponi, Fractal free
  energy landscapes in structural glasses, {\it Nature communications\/} {\bf
  5}, 1 (2014).

\bibitem{wu_2015_2}
B.~Wu, T.~Iwashita, T.~Egami, Anisotropic stress correlations in
  two-dimensional liquids, {\it Physical Review E\/} {\bf 91}, 032301 (2015).

\bibitem{chowdhury_2016}
S.~Chowdhury, S.~Abraham, T.~Hudson, P.~Harrowell, Long range stress
  correlations in the inherent structures of liquids at rest, {\it Journal of
  Chemical Physics\/} {\bf 144}, 124508 (2016).

\bibitem{maier_2017}
M.~Maier, A.~Zippelius, M.~Fuchs, Emergence of long-ranged stress correlations
  at the liquid to glass transition, {\it Physical Review Letters\/} {\bf 119},
  265701 (2017).

\bibitem{lemaitre_2017}
A.~Lema{\^\i}tre, Inherent stress correlations in a quiescent two-dimensional
  liquid: Static analysis including finite-size effects, {\it Physical Review
  E\/} {\bf 96}, 052101 (2017).

\bibitem{lemaitre_2018}
A.~Lema{\^\i}tre, Stress correlations in glasses, {\it Journal of Chemical
  Physics\/} {\bf 149}, 104107 (2018).

\bibitem{zheng_2018_2}
J.~Zheng, A.~Sun, Y.~Wang, J.~Zhang, Energy Fluctuations in Slowly Sheared
  Granular Materials, {\it Physical Review Letters\/} {\bf 121}, 248001 (2018).

\bibitem{degiuli_2018_2}
E.~DeGiuli, Field theory for amorphous solids, {\it Physical Review Letters\/}
  {\bf 121}, 118001 (2018).

\bibitem{lerner_2020}
E.~Lerner, Simple argument for emergent anisotropic stress correlations in
  disordered solids, {\it Journal of Chemical Physics\/} {\bf 153}, 216101
  (2020).

\bibitem{harris1974effect}
A.~B. Harris, Effect of random defects on the critical behaviour of Ising
  models, {\it Journal of Physics C: Solid State Physics\/} {\bf 7}, 1671
  (1974).

\bibitem{weinrib1984long}
A.~Weinrib, Long-range correlated percolation, {\it Physical Review B\/} {\bf
  29}, 387 (1984).

\bibitem{makse2000tracer}
H.~A. Makse, J.~S. Andrade~Jr, H.~E. Stanley, Tracer dispersion in a
  percolation network with spatial correlations, {\it Physical Review E\/} {\bf
  61}, 583 (2000).

\bibitem{marinov2006percolation}
V.~I. Marinov, J.~L. Lebowitz, Percolation in the harmonic crystal and voter
  model in three dimensions, {\it Physical Review E\/} {\bf 74}, 031120 (2006).

\bibitem{prakash1992structural}
S.~Prakash, S.~Havlin, M.~Schwartz, H.~E. Stanley, Structural and dynamical
  properties of long-range correlated percolation, {\it Physical Review A\/}
  {\bf 46}, R1724 (1992).

\bibitem{abete2004percolation}
T.~Abete, A.~De~Candia, D.~Lairez, A.~Coniglio, Percolation model for enzyme
  gel degradation, {\it Physical review letters\/} {\bf 93}, 228301 (2004).

\bibitem{saberi2010geometrical}
A.~A. Saberi, Geometrical phase transition on WO 3 surface, {\it Applied
  Physics Letters\/} {\bf 97}, 154102 (2010).

\bibitem{schmittbuhl1993percolation}
J.~Schmittbuhl, J.-P. Vilotte, S.~Roux, Percolation through self-affine
  surfaces, {\it Journal of Physics A: Mathematical and General\/} {\bf 26},
  6115 (1993).

\bibitem{schrenk2013percolation}
K.~Schrenk, {\it et~al.\/}, Percolation with long-range correlated disorder,
  {\it Physical Review E\/} {\bf 88}, 052102 (2013).

\bibitem{bernard2007inverse}
D.~Bernard, G.~Boffetta, A.~Celani, G.~Falkovich, Inverse turbulent cascades
  and conformally invariant curves, {\it Physical review letters\/} {\bf 98},
  024501 (2007).

\bibitem{he_1985}
H.~He, M.~F. Thorpe, Elastic properties of glasses, {\it Phys. Rev. Lett.\/}
  {\bf 54}, 2107 (1985).

\bibitem{narayanan_1996}
R.~A. Narayanan, S.~Asokan, A.~Kumar, Evidence concerning the effect of
  topology on electrical switching in chalcogenide network glasses, {\it Phys.
  Rev. B\/} {\bf 54}, 4413 (1996).

\bibitem{uebbing_1996}
B.~Uebbing, A.~J. Sievers, Role of Network Topology on the Vibrational Lifetime
  of an H 2 O Molecule in the Ge-As-Se Glass Series, {\it Phys. Rev. Lett.\/}
  {\bf 76}, 932 (1996).

\bibitem{phillips_1979}
J.~C. Phillips, Topology of covalent non-crystalline solids I: Short-range
  order in chalcogenide alloys, {\it Journal of non-crystalline solids\/} {\bf
  34}, 153 (1979).

\bibitem{peshkov_2021}
A.~Peshkov, S.~Teitel, Critical scaling of compression-driven jamming of
  athermal frictionless spheres in suspension, {\it Phys. Rev. E\/} {\bf 103},
  L040901 (2021).

\bibitem{jin_2021}
Y.~Jin, H.~Yoshino, A jamming plane of sphere packings, {\it Proc. Nat. Acad.
  Sci. USA\/} {\bf 118} (2021).

\bibitem{allen2017computer}
M.~P. Allen, D.~J. Tildesley, {\it Computer simulation of liquids\/} (Oxford
  university press, 2017).

\end{thebibliography}

\bibliographystyle{Science_updated}

\end{document}